\documentclass[oldversion]{aa}
\usepackage{graphicx}
\usepackage[authoryear]{natbib}
\usepackage{txfonts}
\begin{document}
\title{Variable pulse profiles of Her X-1 repeating  with the same \\ 
irregular 35\,d clock as the turn-ons}
%   \subtitle{I. Overviewing the $\kappa$-mechanism}

\author{R.~Staubert\inst{1}, 
D.~Klochkov\inst{1}, D.~Vasco\inst{1},
K.~Postnov\inst{2}, N.~Shakura\inst{2}, 
J.~Wilms\inst{3}, R.E.~Rothschild\inst{4}}

\offprints{staubert@astro.uni-tuebingen.de}

\institute{
	Institut f\"ur Astronomie und Astrophysik, Universit\"at T\"ubingen,
	Sand 1, D-72076 T\"ubingen, Germany
\and
	Sternberg Astronomical Institute, 13 Universitetskii pr., 119992 Moscow, Russia
\and
        Dr.\ Remeis-Sternwarte, Astronomisches Institut der
	Universit\"at Erlangen-N\"urnberg, Sternwartstr. 7, 96049 Bamberg, Germany
\and
        Center for Astrophysics and Space Sciences, University of
        California at San Diego, La Jolla, CA 92093-0424, USA
}

%   \email{staubert@astro.uni-tuebingen.de}

\date{Received 31 Aug 2012; accepted 05 Dec 2012}
\authorrunning{Staubert et. al.}
\titlerunning{Variable pulse profiles  in Her X-1}

% \abstract{}{}{}{}{} 
% 5 {} token are mandatory

  \abstract
 {The accreting X-ray pulsar Her X-1 shows two types of long-term variations, both with periods 
 of $\sim$35\,days: 
1) \textsl{Turn-on cycles, a modulation of the flux}, with a ten-day long \textsl{Main-On} and a 
five-day long \textsl{Short-On}, separated by two \textsl{Off}-states, and 
2) a systematic \textsl{variation in the shape of the 1.24\,s pulse profile}. While there is general 
consensus that the flux modulation is due to variable shading of the X-ray emitting regions on 
the surface of the neutron star by the precessing accretion disk, the physical reason for the 
variation in the pulse profiles has remained controversial. Following the suggestion 
that \textsl{free precession} of the neutron star may be responsible for the variation in the pulse 
profiles, we developed a physical model of strong feedback interaction between the neutron star 
and the accretion disk in order to explain the seemingly identical values for the periods of the two 
types of variations, which were found to be in basic synchronization. In a deep analysis of pulse profiles 
observed by several different satellites over the last three decades we now find that the clock behind 
the pulse profile variations shows exactly the same erratic behavior as the turn-on clock, even on 
short time scales (a few 35\,d cycles), suggesting that there may in fact be \textsl{only one 35\,d clock} 
in the system. If this is true, it raises serious questions with respect to the idea of free precession of the 
neutron star, namely how the neutron star can change its precessional period every few years by up 
to 2.5\% and how the feedback can be so strong, such that these changes can be transmitted to the 
accretion disk on rather short time scales.
}


   \maketitle

\section{Introduction}\footnote{A matrix representing the pulse profile 
template is only available in electronic form at the CDS via anonymous ftp to
\texttt{cdsarc.u-strasbg.fr (130.79.128.5)} or via \texttt{http://cdsarc.u-strasbg.fr/viz-bin/qcat?J/A+A}}

%Fig. 1-new
\begin{figure}
\begin{center}
\includegraphics[width=0.50\textwidth]{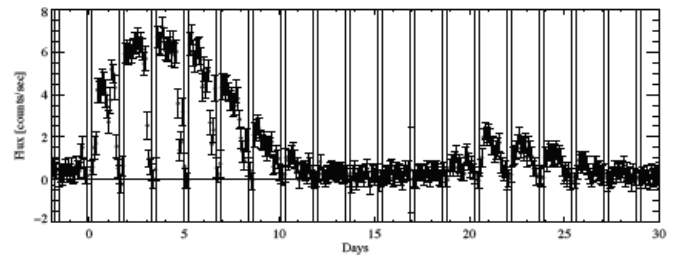} 
\vspace{-3mm}
\caption{Average 35\,d flux profile of Her X-1, generated by accumulating
light curves observed by \textsl{RXTE}/ASM for 35\,d turn-ons around binary
phase 0.2 (Fig.~3 of Klochkov et al. 2006). The vertical lines indicate the binary
eclipses.}
\end{center}
\label{35d_mod}
\end{figure}

The binary X-ray pulsar Her~X-1 shows a number of periodic modulations of its
X-ray flux: the 1.24\,s pulse period, the 1.70\,d orbital period (through eclipses and 
the Doppler modulation of the pulse period), a 1.62\,d dip period, and a 35\,d 
super-orbital period. The last period is observed first through an on-off cycle with a 
10\,d \textsl{Main-On} and a 5\,d \textsl{Short-On} (Fig. \ref{35d_mod}), separated by two 10\,d 
\textsl{Off}-states \citep{Tananbaum_etal72}, and second through a reproduced change 
in the shape of the 1.24\,s pulse profile \citep{Truemper_etal86, Deeter_etal98, Scott_etal00}. 
With respect to these modulations, \citet{Staubert_etal09} argued that there may be two 
$\sim$35\,d clocks in the system which are generally synchronized by strong feedback.

The 35-day modulation of the X-ray flux is generally explained by the precession of 
the accretion disk, which quasi-regularly blocks the line of sight to the X-ray emitting regions 
near the magnetic poles of the neutron star \citep{GerendBoynton_76, SchandlMeyer_94}. 
The resulting clock, however, is irregular, showing deviations from regularity of up to $\pm 10$\,d. 
This is generally demonstrated by the (O--C) (observed-minus-calculated) diagram 
(see Fig~2), plotting the difference between the observed turn-ons and the calculated turn-ons, 
under the assumption of a constant period (e.g. $20.5 \times P_\mathrm{1.7}$ = 34.850\,d; 
see \citealt{Staubert_etal83,StillBoyd_04,Staubert_etal07}). This diagram has very interesting properties
that warrant further attention: 1) a quasi-periodic modulation with a period of $\sim5$\,yr, corresponding 
to a repeated change in the mean precessional period by a few percent, 2) an apparent correlation with 
the appearance of the \textsl{Anomalous Lows}, which are thought to be 
episodes of low tilt \citep{Staubert_etal06a} or higher twist \citep{LeahyDupui_10}
of the accretion disk leading to a blocking of the line of sight to the X-ray 
source for some period of time (ranging from a few days to a few years); 3) a possible long-term
modulation with $\sim$15.5\,yr \citep{Staubert_etal11};
4) substructure with a quasi-periodicity of $\sim2.5$\,yrs in the \textsl{UHURU} era; 5) a correlation
with the X-ray luminosity \citep{Klochkov_etal09}; and most remarkably, 6) an intriguing correlation 
with the history of the pulse-period evolution \citep{Staubert_etal06a}. The last two properties 
require strong physical coupling between the precession of the outer edge of the 
accretion disk and the accretion torques acting on the neutron star when the material at
the inner edge of the accretion disk interacts with the magnetosphere of the neutron star.
These facts lend important support for the feedback scenario described in \citet{Staubert_etal09}.

With regard to the systematic variation in shape of the X-ray pulse profiles, the situation
is considerably less clear. \citet{Truemper_etal86} (based on observations of Her X-1 by
\textsl{EXOSAT}) had proposed that free precession may be the reason for the changing
pulse profiles: due to the variation in the angle of the line of sight with respect to the beamed
emission from the surface of the spinning neutron star. 

%Fig. 2-new
\begin{figure}
\begin{center}
\resizebox{1.15\hsize}{!}{\includegraphics[angle=-90]{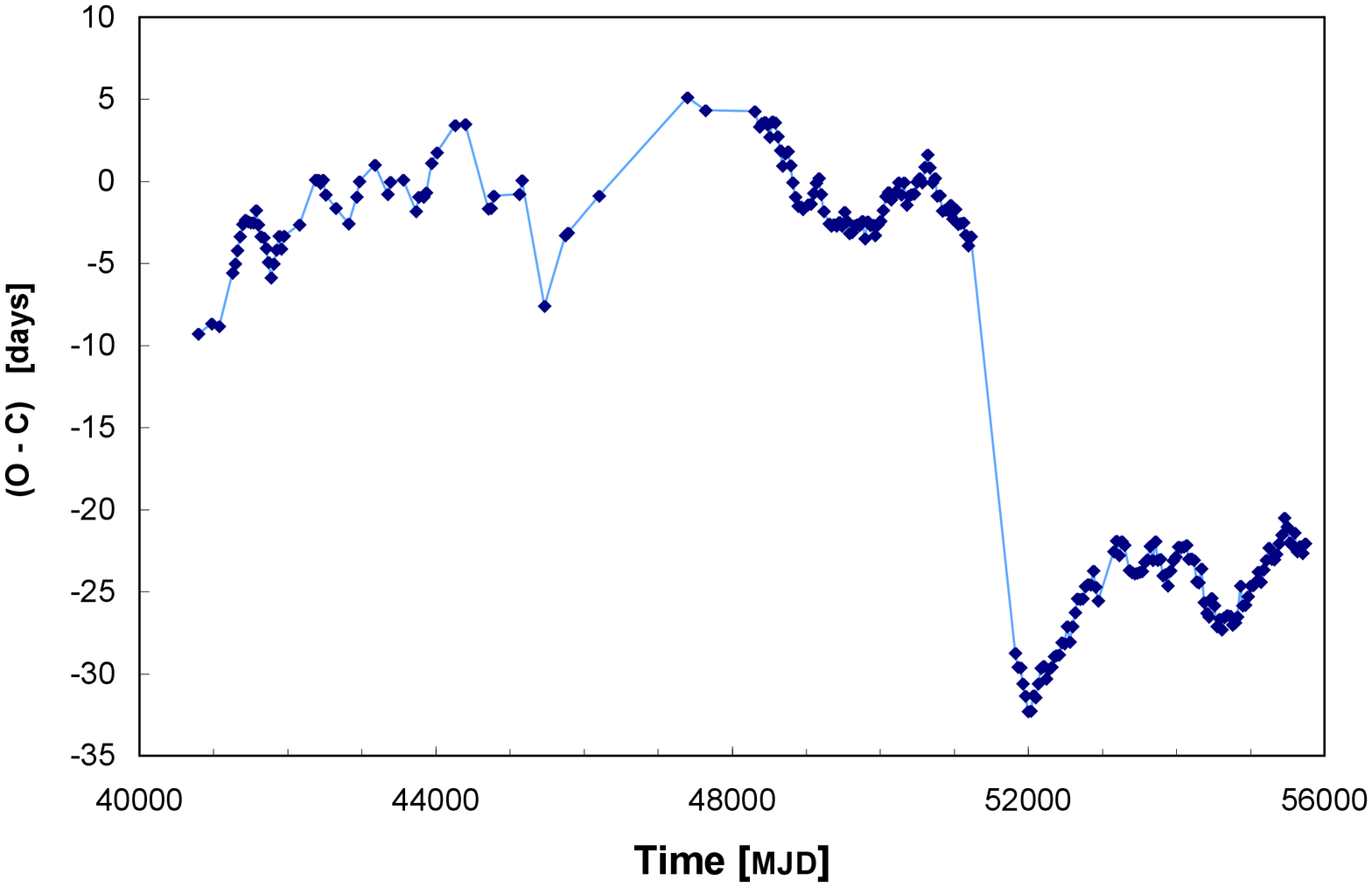}}
\hfill
\vspace{-15mm}
\resizebox{1.15\hsize}{!}{\includegraphics[angle=90]{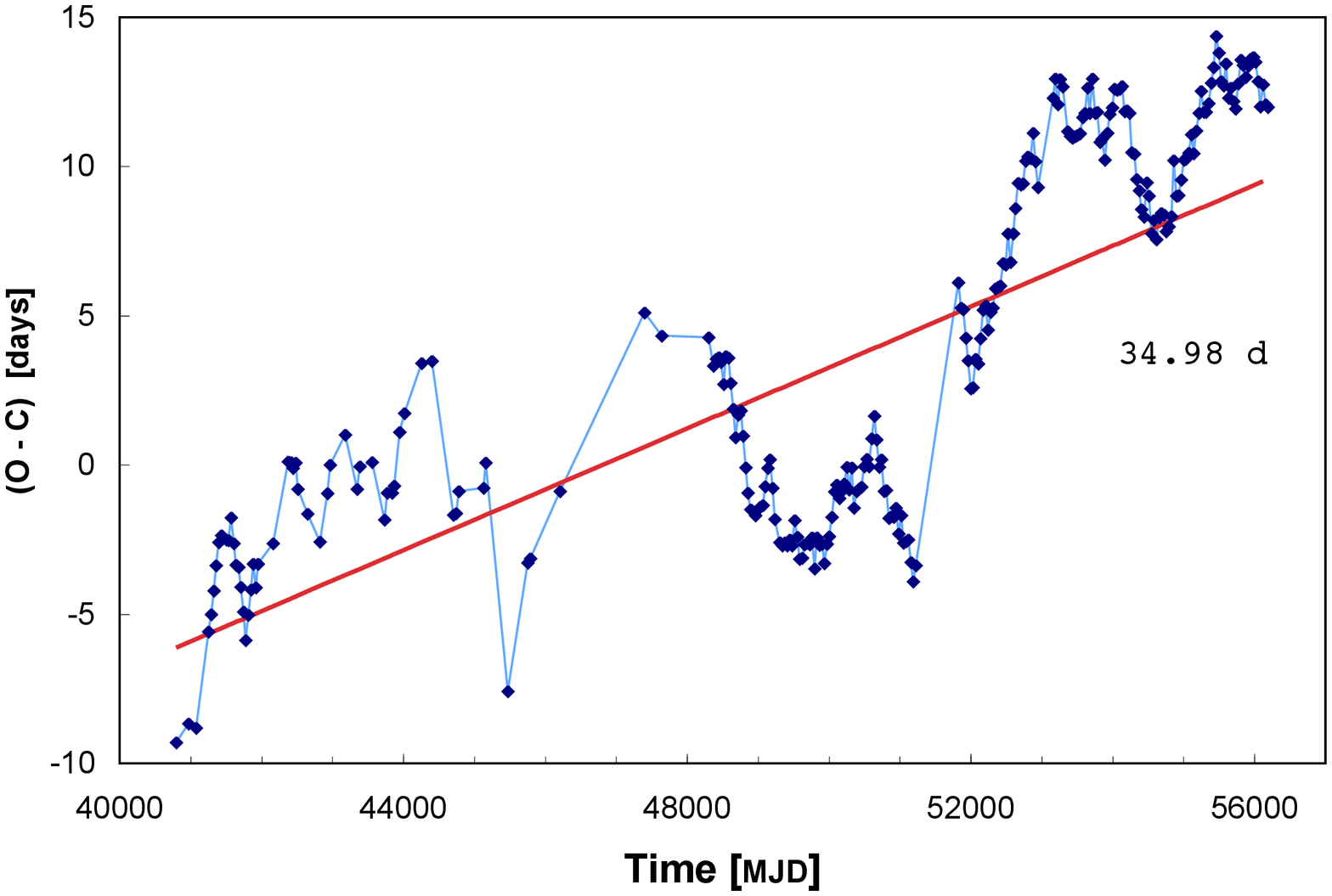}}
\vspace{-5mm}
\caption{Turn-on history of Her~X-1: the ($O-C$) diagram (using 
20.5$\times$P$_\mathrm{orb}$ as the period to calculate the expected turn-ons).
\textsl{Upper}: The Anomalous Low around MJD 51500 (AL3) is assumed to last for 
eighteen 35\,d cycles. 
\textsl{Lower}: The same assuming that there were only seventeen 35\,d cycles
(see \citealt{Staubert_etal09}). The solid line represents the linear best fit to all data, 
defining a mean period of 34.88\,d (which may be used for a rough ephemeris of turn-ons).}
\end{center}
 \label{O-C_basic}
\end{figure}

Free precession may appear as a fundamental physical property of rigid 
non-spherical spinning bodies. The simplest case is a spheroid with 
some small oblateness (a ``two-axial" body) in which the axis of angular velocity 
is not aligned with any principle axis of inertia (e.g., \citealt{Sommerfeld_Klein_97}). 
It has been suggested as the underlying  reason
for the long-period variations, both in timing and spectral properties, 
observed in several neutron stars  (\citealt{JonesAnderson_01, Cutler_etal03, LinkEpstein_01, 
Haberl_etal06}). The candidate objects are mostly radio pulsars (including the Crab 
and Vela pulsars), the isolated X-ray pulsar $RX J0720.4-3125$
\citep{Haberl_etal06} and the accreting binary X-ray pulsar Her~X-1.
The existence of free precession in  neutron stars and its consequences for
our understanding of the physics of the interior of neutron stars is extensively 
discussed in the literature (\citealt{AndersonItoh_75,Shaham_77,
AlparOegelman_87,Sedrakian_etal99,Wasserman_03,LevinDangelo_04,
Link_07}). Recently, \citet{Link_07} has emphasized that the question of the
reality of free precession in neutron stars has strong implications for our
understanding of the properties of matter at supra-nuclear densities.

The idea of free precession in Her~X-1 has been taken up by several authors (e.g. 
\citealt{Shakura_etal98,Ketsaris_etal00,Postnov_04,Staubert_etal07,Postnov_etal12}). 
Recently, \citet{Postnov_etal12} have successfully modeled 9--13\,keV pulse profiles 
observed by \textsl{RXTE} under a particular set of assumptions.
However, the idea has also been questioned on various grounds (e.g., 
\citealt{BisKog_89,Soong_etal87,Scott_etal00,Staubert_etal10b}) and alternative models for 
the generation of variable pulse profiles, mostly involving the inner edge of the accretion disk 
and/or the accretion column have been proposed 
(e.g. \citealt{Petterson_etal91,Scott_etal00,Leahy_04}).
 
Here we present the results of a model independent study of the variation in pulse
profiles in comparison to the variation in the 35\,d flux modulation. We find, that the two
ways to count 35\,d cycles - observing the turn-ons and observing the change in pulse
profile shape - track each other perfectly, following one and the same irregular clock.
First results of this study were presented by \citealt{Staubert_etal10b}.
It appears, that we cannot talk about two different clocks anymore, but we have to 
conclude that \textsl{there is really only one clock in the system}. This result shines new light 
onto the question of how the pulse profiles are generated and what the consequences 
are for the concept of neutron star free precession in Her~X-1.

%Fig. 3-new
\begin{figure*}[t]
      \includegraphics[angle=-90,width=0.50\textwidth]{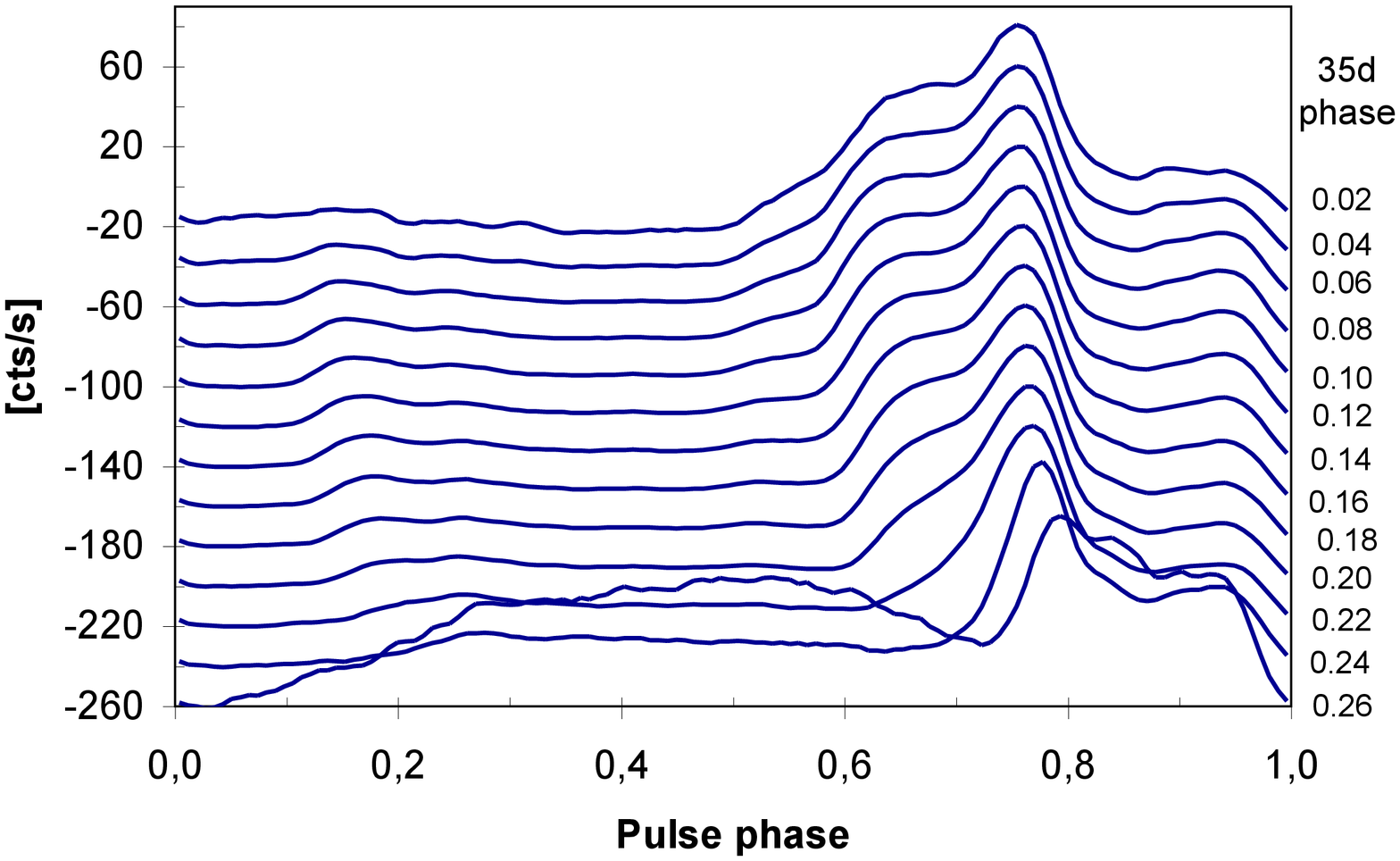}
     \hfill
     \includegraphics[angle=-90,width=0.50\textwidth]{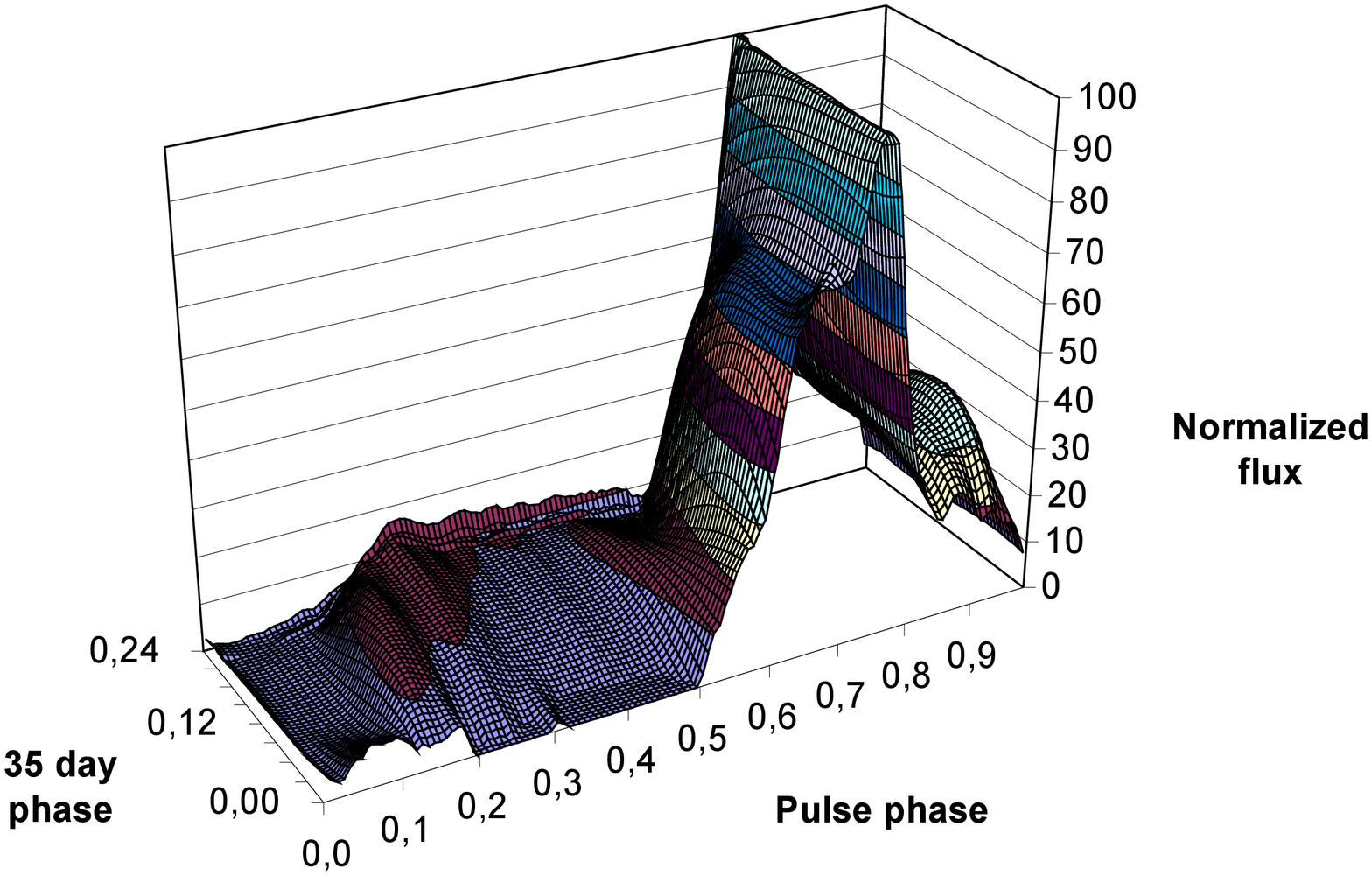}
   \caption{Representation of the template which describes \textsl{Main-On} pulse profiles of Her~X-1 
   in the 9--13\,keV range as function of 35\,d phase. 
  \textsl{Left}: Pulse profiles for every 0.02 in 35\,d phase.
   \textsl{Right}: 3D pulse representation for the 35\,d phase range 0.00 to 0.24.
    The resolution is 128 bins in pulse phase and 48 bins in 35\,d phase (resolution 0.005).}
   \label{compar}
\end{figure*}

\section{35\,d flux modulation and turn-on history}

The 35\,d modulation shows a sharp increase of flux towards the \textsl{Main-On}. It is called 
\textsl{turn-on}, and defines 35\,d phase zero; it generally occurs either around binary phases 0.2 or 
0.7 (\citealt{Giacconi_etal73,LevineJernigan_82}; see, however, \citealt{LeahyIgna_10}).
Fig. \ref{35d_mod} shows the mean flux profile for phase 0.2 turn-ons as observed by 
\textsl{RXTE}/ASM \citep{Klochkov_etal06}.  
The 35\,d turn-on clock is fairly irregular, allowing the length of an individual cycle to be either
20.0$\times$P$_\mathrm{orb}$, 20.5$\times$P$_\mathrm{orb}$, or 21.0$\times$P$_\mathrm{orb}$ 
\citep{Staubert_etal83}
(with a small fraction of cases showing longer or shorter cycles). 
Adopting P$_{35}$ = 20.5$\times$P$_\mathrm{orb}$ (= 34.85\,d) as the mean ephemeris period, the 
turn-on history can be described by the ($O-C$) diagram.
Fig.~2 (upper and lower) shows ($O-C$) since the discovery of 
Her~X-1 until today as a function of 
time.\footnote{The turn-on times listed by \citet{LeahyIgna_10} for a restricted period of time are
largely consistent with our values.}
The two versions of the diagram (upper and lower)
differ in the number of 35\,d cycles assumed to have occurred during the 602 day long 
Anomalous Low (AL3), which is centered at $\sim$\,MJD~51500. During this gap, the observed X-ray flux 
was very low because of the blocking by the low inclination accretion disk 
(optical observations indicate that X-rays are still generated at the neutron star surface, but not seen
by a distant observer).
For the interpretation of Fig.~2
%Fig.~\ref{O-C_basic} 
we refer to \citet{Staubert_etal09}, in the following called the "Two-clocks paper".
In Fig.~2-upper
%Fig.~\ref{O-C_basic}-upper 
the  gap corresponds to 18 cycles with a (short) mean duration of 
19.7$\times$P$_\mathrm{orb}$, which is probably the correct physical interpretation for a continued 
precession of the accretion disk.\footnote{A detailed analysis of the \textsl{RXTE}/ASM data of AL3
is in progress.}
In Fig.~2-lower
%Fig.~\ref{O-C_basic}-lower 
the gap corresponds to 17 cycles with a 
(long) mean duration of 20.8$\times$P$_\mathrm{orb}$, which can be associated with a 
semi-regular, long-term clock with a mean period of 20.58$\times$P$_\mathrm{orb}$ = 34.98\,d. In the 
"Two-clocks paper" \citep{Staubert_etal09} this period is associated with an underlying clock, viewed to 
be rather regular, namely (free) precession of the neutron star, assumed to be 
responsible for the observed periodic variation in the shape of the pulse profiles. In this model, 
it is assumed, that the precession of the neutron star is the \textsl{master clock}
and that the precession of the accretion disk is evidently locked to that of the neutron star by 
existing strong physical feedback in the system, (nearly) synchronizing the periods of the 
"two clocks".

%\vspace{-6mm}
\section{Pulse profile variations}

In observations of Her~X-1 by \textsl{EXOSAT} in 1984, \citet{Truemper_etal86}
discovered that the 1.24\,s X-ray pulse profiles vary in shape as a function of 
the phase of the 35\,d flux modulation. Observations by \textsl{Ginga} in 1989
\citep{Scott_etal00}, by \textsl{HEAO-1} \citep{Soong_etal90}
and by \textsl{RXTE} starting in 1996 \citep{Staubert_etal10a, Staubert_etal10b},
confirmed these findings and added a wealth of detailed information on the 
combined pulse shape and spectral evolution of the pulsar's beamed emission. 
\citet{Truemper_etal86} had suggested that the systematic variation in pulse shape is 
due to free precession of the neutron star: the viewing angle towards the X-ray emitting 
regions of the neutron star varies with the phase of the neutron star precession.
\citet{Shakura_etal98} applied a model of a precessing triaxial shape to pulse
profiles of Her~X-1 observed by \textsl{HEAO-1}, and \citet{Ketsaris_etal00} did so
for profiles observed by \textsl{RXTE}/PCA. 
Using all observations by \textsl{RXTE} from 1996 until 2005 we have verified
that the shape of the pulse profiles is reproduced every $\sim35$ days. 
A careful timing analysis was performed of all archived \textsl{RXTE} data on Her X-1 
and pulse profiles were generated by folding with the measured pulse periods.  
Fig. \ref{compar} shows two representations of the final template: \textsl{left}:
a set of pulse profiles for 35\,d phases 0.02 to 0.26 (every 0.02), and \textsl{right}:
a 3D-plot for 35\,d phases 0.02 to 0.24 with a resolution of 0.005 (the resolution in
pulse phase is 128 bins for 0.0 to 1.0). The variation in pulse shape is evident,
most pronounced is the gradual disappearance of the left shoulder
of the main peak.
This systematic analysis had led to the development of the "Two-Clocks-Model" 
\citep{Staubert_etal09} and the successful modeling of the observed pulse profiles 
by a model of point- and ring-like emission from the polar caps of a neutron star 
with an offset-dipole field under the assumption of free neutron star precession 
\citep{Postnov_04, Postnov_etal12}.

\citet{Staubert_etal10b} presented a model-independent investigation of the 
periodic pulse profile variations based upon constructing a template of those variations 
for the Her~X-1 \textsl{Main-On} for photon energies 9--13\,keV.
A set of observations was selected, providing good 
coverage of the 35\,d phase range -0.05 to 0.15. This template contained flux normalized
pulse profiles for every 0.01 in phase. Any 9--13\,keV pulse profile observed during
a \textsl{Main-On} could then be compared to this template and the 35\,d phase be determined 
(by $\chi^{2}_\mathrm{min}$ minimization). For the \textsl{RXTE} data, we found that this is 
generally possible 
to an accuracy of  $\pm$0.02 in phase. First results of using this template with profiles
observed by \textsl{Ginga}, \textsl{RXTE} and \textsl{INTEGRAL} were presented
by \citet{Staubert_etal10a,Staubert_etal10c}, showing that the turn-on history and the history of 
variations in pulse profiles appeared to be strictly parallel, implying that the neutron star
precession (if it is indeed the reason behind the pulse profile variations) is synchronized
to the precession of the accretion disk. This raised the important question about how the neutron 
star would be able to significantly change its precessional period every few years by up to
2.5\% and how such changes are relayed to the accretion disk on rather short time scales.

We have since refined this method by constructing an extended template from \textsl{RXTE}/PCA
observations of two 35\,d cycles, namely cycles no. 313 (Dec 2001) and no. 323 (Nov 2002).
Together, the data from the observation of these two cycles provide a very well sampled coverage 
of a complete \textsl{Main-On} in the 35\,d phase range 0.00 to 0.24 (Fig.~\ref{lcs}).
The details of the construction of this template and its exact definition are given in the \textsl{Appendix}. 
An earlier version of the template (with less resolution) is described in \citet{Staubert_etal10b}. 
A 3D-representation of the new template (being a matrix of 128 bins in pulse phase by 121 bins in 35\,d 
phase) is given in Fig. \ref{compar}-right. 

%------------ TABLE 1 -----------------
\begin{table*}
\vspace*{2mm}
\caption{Details of \textsl{RXTE} observations of Her~X-1 used for the pulse profile analysis.} 
\label{obs_table}
%\centering
\begin{tabular}{l l l l l l l l l l}
\hline\hline
Cycle	 &	Satellite  	   &	Observed	&		& PP phase	&	       & Calculated	  &  (O - C)           &  (O - C) of	& Diff. btw. Turn-On\\
no.$^{1}$	&			   &	Turn-On	&	 +/-   & zero $^{2}$	&	 +/-  & Turn-On$^{3}$ &  of turn-on	    &  PP phase zero	& and PP phase zero \\
		&			   &	(MJD)	&   (days) &    (MJD) 	&  (days)	&	(MJD)	   &	(days)	    &	(days)	&  (sigma) \\
		&			   &			&		&			&		&			   &			    &		        &	 \\
181		&	Ginga	   &	47642.20	&	0.20	&	47642.13	&	0.11	&     47637.87	   &	4.33		    &	4.26		&	0.31 \\
252		&	RXTE	   &	50111.80	&	0.40	&	50111.43	&	0.40	&	50112.46	   &	-0.66	    &	-1.03	&	0.65 \\
257		&	RXTE	   &	50285.90	&	0.20	&	50285.46	&	0.20	&	50286.73	   &	-0.83	    &	-1.27	&	1.56 \\
259		&	RXTE	   &	50356.00	&	0.60	&	50354.53	&	0.20	&	50356.43	   &	-0.43	    &	-1.91	&	2.33 \\
269		&	RXTE	   &	50704.90	&	0.20	&	50704.27	&	0.20	&	50704.97	   &	-0.07	    &	-0.70	&	2.24 \\
271		&	RXTE	   &	50773.80	&	0.30	&	50771.17	&	0.80	&	50774.68	   &	-0.88	    &	-3.51	&	3.08 \\
301		&	Beppo/SAX  &	51826.40	&	0.50 &	51825.83	&	0.20	&	51820.28	   &	6.12	 	    &	5.55		&	1.06 \\
303		&	RXTE	   &	51895.20	&	0.50	&	51894.58	&	0.22	&	51889.99	   &	5.21		    &	4.59		&	1.14 \\
304		&	RXTE	   &	51929.10	&	0.50	&	51929.43	&	0.50	&	51924.84	   &	4.26		    &	4.59		&	-0.46 \\
307		&	RXTE	   &	52032.00	&	0.50	&	52031.37	&	0.21	&	52029.40	   &	2.60		    &	1.97		&	1.16 \\
308		&	RXTE	   &	52067.80	&	0.50	&	52066.46	&	0.25	&	52064.25	   &	3.55		    &	2.21		&	2.40 \\
313		&	RXTE	   &	52243.05	&	0.50	&	52243.04	&	0.06	&	52238.52	   &	4.53		    &	4.52		&	0.02 \\
319		&	RXTE	   &	52454.40	&	0.50	&	52453.89	&	0.50	&	52447.64	   &	6.76		    &	6.25		&	0.72 \\
320		&	RXTE	   &	52489.20	&	0.50	&	52488.73	&	0.50	&	52482.49	   &	6.71		    &	6.23		&	0.67 \\
323		&	RXTE	   &	52594.66	&	0.50	&	52594.65	&	0.02	&	52587.05	   &	7.61		    &	7.60	        &	0.02 \\
324		&	RXTE	   &	52630.50	&	0.50	&	52629.48	&	0.50	&	52621.91	   &	8.59		    &	7.57		&	1.44 \\
340		&	RXTE	   &	53192.50	&	1.00	&	53192.03	&	0.80	&	53179.56	   &	12.94	    &	12.47	&	0.37 \\
343		&	RXTE	   &	53296.80	&	0.50	&	53297.17	&	0.80	&	53284.12	   &	12.68	    &	13.04	&	-0.39 \\
353		&	Suzaku 05   &	53645.30	&	0.20	&	53644.43	&	0.80	&	53632.66	   &	12.64	    &	11.77	&	1.06 \\
358		&	Suzaku 06   &	53817.75	&	0.20	&	53816.37	&	0.80	&	53806.92	   &	10.83	    &	9.45		&	1.67 \\
373		&	INTEGRAL  &	54341.00	&	0.50	&	54339.06	&	1.20	&	54329.73	   &	11.27	    &	9.33		&	1.50 \\
\hline
\end{tabular} 
\vspace*{1mm} \\
$^{1}$ The numbering of 35\,d cycles follows the convention introduced by \citet{Staubert_etal83}:  $(O - C)$ = 0 for cycle no. 31 with turn-on \\
near JD 2442410. See, however, \citealt{Staubert_etal09} for the ambiguity of counting after the long \textsl{Anomalous Low} in 1999/2000 (AL3). \\
 $^{2}$ "PP phase zero'' stands for: "pulse profile phase zero" (see text). \\
$^{3}$ again: 20.5$\times$P$_\mathrm{orb}$ is assumed as the period to calculate the expected turn-ons.
\label{obs}
\end{table*}
%------------------------------------

%Fig. 4
\begin{figure}
  \vspace*{-3mm}
  \includegraphics[angle=-90,width=0.57\textwidth]{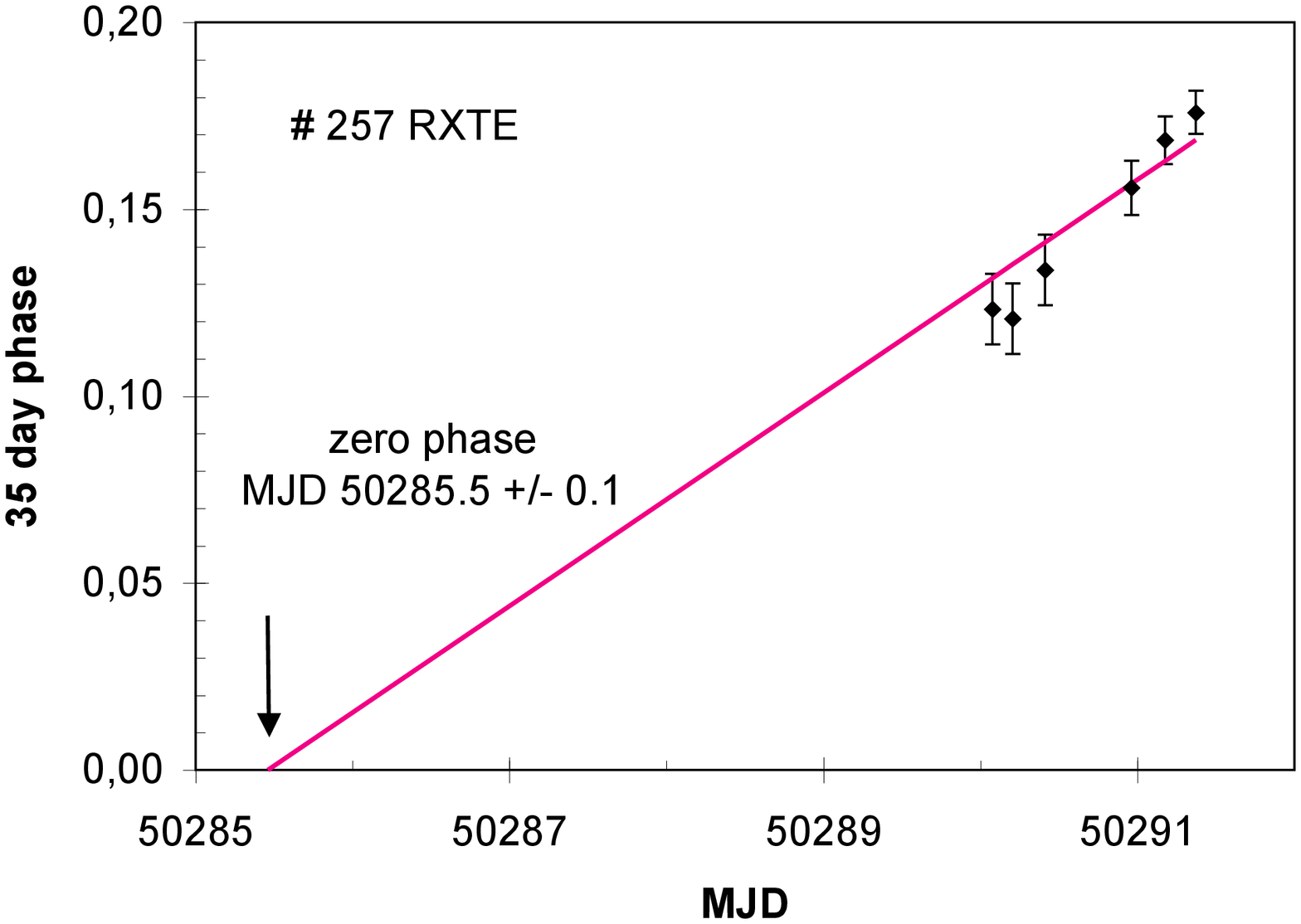}
  \vspace*{-8mm}
  \caption{Example for finding "pulse profile phase-zero" for the case of cycle 257 
                Main-On, as observed by \textsl{RXTE}/PCA in July 1996.}
   \label{find_phase_zero}
\end{figure}

\section{Results}

\begin{figure*}
  \vspace{-8mm}
\includegraphics[width=0.7\textwidth,angle=-90]{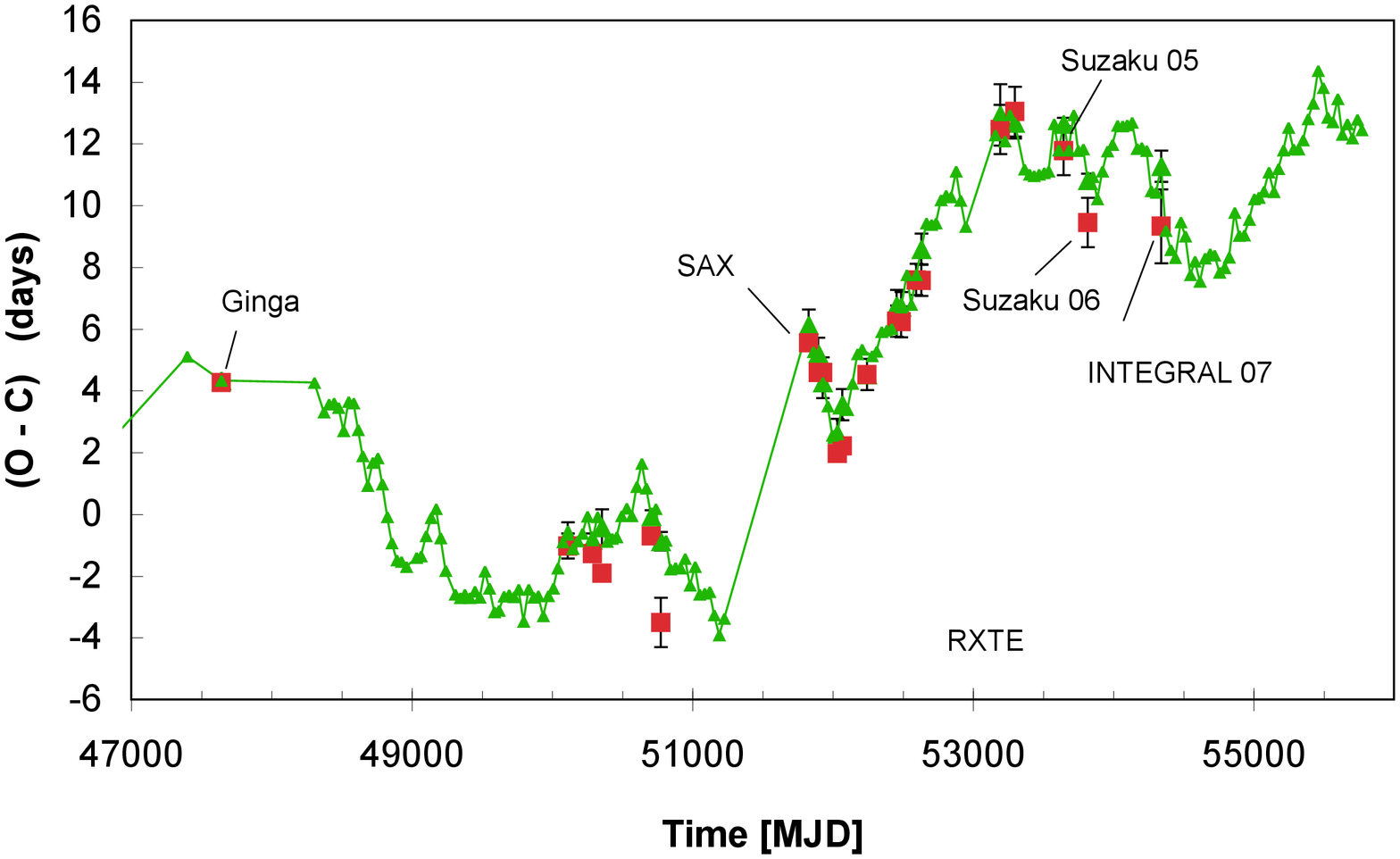} 
  \vspace{-17mm}
  \caption{($O-C$) values in units of days for observed turn-ons (green) and for so far generated
                "pulse profile phase-zero" values from pulse profile fitting (magenta),
                using profiles 9--13\,keV profiles for several Main-Ons. The majority of the 
                 "pulse profile phase-zero" values are from observations by \textsl{RXTE}. The values
                based on observations by \textsl{Ginga}, \textsl{Beppo}/SAX, \textsl{Suzaku} and 
                \textsl{INTEGRAL} are marked individually.}
\label{O-C_final}
\end{figure*}

We compare the characteristics of the two 35\,d modulations (of the flux and of the variation in
pulse profiles) by plotting both into an ($O-C$) diagram (Fig.~\ref{O-C_final}). The underlying
data are summarized in Table~\ref{obs}, together with the 35\,d cycle number and the satellite
which performed the corresponding \textsl{Main-On} observations of Her~X-1.

For the flux modulation, we use the turn-on, also called the "accretion disk phase-zero" and 
plot the ($O-C$) as in Fig.~2.
%Fig.~\ref{O-C_basic}. 
Correspondingly, we determine a "pulse profile phase-zero" for all the \textsl{Main-Ons} for which 
there is at least one observed pulse profile (for most Main-Ons there are several profiles). The 
"pulse profile phase-zero" is defined and calibrated using those cycles which are used to construct 
the template: for cycles 313 and 323 both types of "phase-zero" are identical (by definition). For 
any other cycle "pulse profile phase-zero" is found by determining the 35\,d phases of all available 
pulse profiles by comparison with the template and a subsequent linear extrapolation to phase zero.
Fig.~\ref{find_phase_zero} shows an example for the linear extrapolation to find phase zero. 
For these extrapolations, we use values of the 35\,d period which are correct for the time 
of observation, based on fitting sections of the ($O-C$) diagram by linear functions, as
described in \citet{Staubert_etal11}: to a fairly good approximation, sections (of variable length) 
can be modeled by a respective constant period. Similar attempts of straight line fittings
can be found in \citealt{StillBoyd_04,Jurua_etal11}. The typical change in the precession period 
from section to section is on the order of 1.5\%. At the re-appearance of Her~X-1 after the
\textsl{Anomalous Low} in 1999/2000 (AL3), the precession continued with a rather short period
until around MJD~52000 (see Fig. \ref{O-C_final}), at which time a rapid increase by 2.5\% 
occurred, on a time scale comparable to two 35\,d cycles. We note that within the entire 
($O-C$) data base, the largest change (of $\sim$4\%) is found in the early data from \textsl{UHURU}, 
comparing the initial upward trend in ($O-C$) to the following downward trend (see Fig.~2).

Using pulse profiles of ten other \textsl{Main-On} cycles of Her~X-1 observed by \textsl{RXTE} over 
the last three decades, plus one from \textsl{Ginga}, one from \textsl{Beppo}/SAX, two from 
\textsl{Suzaku} and one from \textsl{INTEGRAL}, we find that these systematic variations are very 
stable and reproducible. The constructed template therefore allows one to determine the 35\,d phase 
for any observed \textsl{Main-On} pulse profile in the 9--13\,keV range.

%Fig. 5: 

The observational result in the form of an ($O-C$) diagram is summarized in Fig.~\ref{O-C_final}. 
The small green points (connected by the solid green line) represent the observed turn-on 
times (equal to "accretion disk phase-zero"). The magenta points are the times of "pulse profile 
phase-zero" as determined from the comparison of observed pulse profiles with the pulse profile 
template and subsequent extrapolation to phase zero. For these cycles the corresponding 
green points (turn-ons) are enlarged and shown with uncertainties.
The observational evidence is clear: within statistical uncertainties,
the values for phase-zero as determined by the two different methods are identical
(see the last column in Table~\ref{obs}, which gives the differences between the
two phase-zero values in units of standard deviations).
This means: \textsl{the "pulse profile clock" is just as irregular as the "turn-on clock".
Both clocks appear to be perfectly synchronized}.

We emphasize here, that the above result is completely model independent,
it is obtained using only observational data. However, the latest results from 
the continued effort in modeling the observed pulse profiles by a model 
assuming free precession of the neutron star \citep{Postnov_04,Postnov_etal12} 
leads to the same conclusion. This does suggest the need to abandon
the concept of "two clocks" and to assume the existence of just \textsl{one 
underlying clock} which controls the variations of both 35\,d zero-phases:
that of the turn-ons, and that of pulse profile phase zero. 

Fig.~\ref{O-C_final} demonstrates that the \textsl{Turn-On} clock is
fairly noisy, with additional quasi-periodic variations on a $\sim$5 year
time scale. $(O - C)$ correlates with the appearance of the \textsl{Anomalous
Lows} (AL), and it also strongly correlates with the neutron star's spin period 
\citep{Staubert_etal06a}.

\section{Discussion}

What does the above result mean for the concept of (free) precession of the
neutron star in Her~X-1? We distinguish between two assumptions.
First, assuming the precession of the neutron star does indeed exist and
is responsible for the variation in pulse profiles and(!) for those of the turn-on
times, we would then have to find a physical explanation for two phenomena.

\begin{enumerate}
\item How can the neutron star change its precessional period every few
  years by a few percent?
\item How does the mechanism of synchronization between the neutron star
  precession and the accretion disk precession work? Can the feedback in the
  system really be strong enough to slave the disk, such that it follows on
  very short time scales (a few 35\,d cycles)?
\end{enumerate}

There seem to be no external forces that are strong enough to 
change the precessional period, e.g. by applying a torque to the principle
axis of inertia. 
The only possible origin could be inside the neutron star,
that is, if glitches occur or if the complex physics of the interior of a highly
magnetized neutron star with its crust and liquid core did allow for time
variable phenomena (see also \citealt{Truemper_etal86} for discussion
of possible mechanisms). For the relevance of free precession to our
understanding of matter at supra-nuclear densities, see e.g., \cite{Link_07}.
The period of free precession P$_{pr}$ of a two-axial symmetric body with 
moments of inertia I$_{1}$= I$_{2}$  and I$_{3}$ (the precession takes 
place around I$_{3}$) is related to the spin period P by the Euler equation
\citep{Sommerfeld_Klein_97}: 
P$_{pr}$ = P  I$_{3}$ / ($\Delta$I~cos$\Theta$), with $\Theta$ being
the angle between I$_{3}$ and the total angular momentum.
The period of free precession can change when either $\Delta$I or
$\Theta$ (or both) change. \citet{Postnov_etal12} argue that $\Delta$I
is probably constant, but that a change in $\Theta$ by about one 
degree, caused e.g. by a sudden relaxation of stresses in the
neutron star crust, could lead to a change in precessional period by 
a few percent (assuming $\Theta$ is of the order 50\,degree for Her X-1). 

Equally challenging is the explanation of how the neutron star, if it is indeed 
the one master 35\,d clock in the system, would be able to transmit its 
precessional motion to that of the accretion disk (assuming that our long-term
concept is correct, that the turn-ons are due to the precession of the
accretion disk). We do indeed see strong feedback in the binary system,
which has led \citet{Staubert_etal09} to assume that the precession
of the neutron star may be the master clock in the system, under the
assumption, however, that the neutron star clock would be rather stable
and that the accretion disk would have "a life of its own" providing the
freedom to deviate from the strict clocking of the master - as observed
in the turn-on history. We now see, that the turn-ons and the pulse profiles
vary in strict synchronization, requiring extremely strong feedback.
Again, \citet{Postnov_etal12} show that - under the geometrical conditions 
believed to be realized in Her~X-1 - a sufficiently strong  feedback could 
exist through the dynamical action of the accretion stream on the accretion 
disk, such that synchronization is enforced. 
Certainly, precession of a neutron star in an accreting binary would 
not really be "free" and quantitative model calculations of the interaction 
between the accretion disk and the magnetized neutron star
are clearly needed.

Second, assuming that there is no neutron star precession, we would
associate the precession of the accretion disk with the one
35\,d clock observed in this system. The generally accepted interpretation
of the 35\,d flux modulation is that an \textsl{inclined / warped / counter-precessing
accretion disk} blocks our view to the X-ray source for most of the cycle,
leaving two stretches of $\sim$10\,d and $\sim$5\,d, the \textsl{Main-On}
and \textsl{Short-On}, respectively, with a more or less (\textsl{Short-On})
clear view to the X-ray source. The disk structure may be viewed as a 
continuous series of rings of increasing radius and inclination (from inside
out), shifted against one another in azimuth (angle of lines of nodes), constituting 
the \textsl{twist} or \textsl{warp}. For details of such models and corresponding 
parameters see e.g. 
\citet{SchandlMeyer_94,WijersPringle_99,Scott_etal00,Leahy_02}.
It  is then natural to assume that the disk is also responsible for the 
changing pulse profile shape, with the inner edge, precessing with the
same period as the outer edge (and all other rings), being the prime 
candidate.
\citet{Scott_etal00}, on the basis of \textsl{Ginga} and early \textsl{RXTE} 
observations, had proposed such a model in which the changes in the shape 
and in the spectral appearance of the pulses were qualitatively explained by 
a combination of occultation of the X-ray emitting regions by the
precessing inner edge of the accretion disk and by changes in the accretion 
geometry by a varying relative orientation between the disk and the neutron 
star magnetosphere. 
The model by  \citet{Scott_etal00} 
requires a very small inner radius of the accretion disk of 20 to 40 neutron
star radii, which is incompatible with most estimates of the magnetospheric 
radius (see e.g. the discussion and references in \citealt{Scott_etal00}). 
The magnetospheric radius is usually calculated assuming a spherically symmetric
dipolar magnetic field with a surface strengths of a few times $10^{12}$ Gauss 
(as consistently found from both the observed cyclotron line energy of 40\,keV 
and the interpretation of the observed spin-up/spin-down behavior by standard 
accretion torque theory, e.g. \citealt{GhoshLamb_79,Wang_87,Perna_etal06}).
In addition, we like to point out that the magnetospheric radius in Her~X-1
should not be much smaller than the co-rotation radius, since the source is known
to live close to its equilibrium with respect to spin-up/spin-down. The co-rotation
radius is near 200 neutron star radii.

For completnes, we mention an earlier model which tries to explain the 
variation in the pulse profile shape between 
the \textsl{Main-On}  and the \textsl{Short-On}, as proposed by \citet{Petterson_etal91}: 
here a portion of the inner edge of the warped, counter-precessing 
accretion disk, rotating synchronously to the neutron star, is raised out of the plane
of the disk due to near vertical magnetic pressure. In this model the change in the pulse 
profiles results entirely from the disk motion. This is consistent with our observations
of the variations throughout the \textsl{Main-On} and supports the idea of having only 
one 35\,d clock, namely accretion disk precession. But again, the model is only qualitative. 
A detailed quantitative model for the generation of variable multi-component 
pulse profiles by accretion disk/magnetospheric effects is not available so far.

Apart from the details of how the pulse profiles are generated, the
following basic questions need to be answered: 
\vspace{-1mm}
\begin{enumerate}
\item What causes the initial tilt and warp of the disk and what keeps it
alive over long time scales? In addition to initial conditions (e.g. the relative 
orientation of neutron star spin and orbital angular momentum at birth of the 
X-ray binary), the existence of internal forces on the accretion disk,
like coronal winds \citep{SchandlMeyer_94}, radiation pressure \citep{Pringle_96,
WijersPringle_99}, or the impact of the accretion stream \citep{Shakura_etal99}
have been proposed to produce and maintain the observed configuration. 
\item What causes the variation of the period of precession (by a few percent)
on time scales from a few years to tens of years? - as apparent from
the ($O-C$) diagram (Fig.~2). The accretion disk is subject to a large number
of torques. In addition to the three already mentioned above, (1) coronal wind,
2) radiation pressure, 3) impact of the stream, there are three more: 4) torque
due to the interaction of the neutron star magnetosphere with the inner edge
of the accretion disk, 5) tidal forces by the binary components, and 5) the
internal viscosity of the disk. A careful analysis of the overall balance of these
torques is needed. But it is evident, considering the various feedback processes 
working in the binary, that both, the whole system and the accretion disk, 
live in a "delicate equilibrium", which allows for deviations from strict regularity
(see the discussion in \citealt{Staubert_etal09}).
It may even be possible that the observed $\sim$5\,yr modulation (and its overtones 
of $\sim$10\,yr and $\sim$15\,yr) represent a natural ringing frequency of
a system of several coupled physical components \citep{Staubert_etal09}.
\item Finally, we can ask whether a system described above could be
secularly stable. Given that the total time span of the existing data base
is short ($\sim40$\,yrs), the observations can not answer the question.
For time scales $<40$\,yrs the ($O-C$) diagram (in the form of Fig.~2-lower) 
suggests that there is a mean precession period which could be stable, and
there seems to be a "back-driving force" which brings the ($O-C$) always back
to the center line, possibly representing a long-term period. 

There is, however, evidence that the accretion disk precession may
not be stable, or at least - should a long-term stability indeed exist - there can
be temporary deviations from stability.
The upper part of  Fig.~2 shows a dramatic step in ($O-C$) around 
MJD~51500 (AL3). We believe (see footnote 2 and \citealt{Staubert_etal09}) 
that this is the correct description of the physical behavior of the accretion disk, 
namely that the precessional period of the accretion disk during AL3 and its immediate 
vicinity was low. So, over the time of this \textsl{Anomalous Low} the accretion disk 
has performed "one extra precessional cycle" as compared to the more regular clock 
(Fig.~2-lower), which in the "Two-clocks model" of  \citet{Staubert_etal09} is associated 
with the precession of the neutron star.  AL3  is a very special event in the long-term 
evolution of the system and it is not clear whether the notion that the system found back to 
the usual (albeit non-perfect) regularity after the AL is an adequate description.
A comprehensive study is needed about the question of stability, which is beyond the scope 
of this contribution. If it turns out that the system can not be stable on long time
scales and that an external clock is needed, we would see two possibilities:
1) Either: free precession of the neutron star, as discussed earlier. This would 
also answer the question of how the disk remains in phase with its 35\,d clock 
during long \textsl{Anomalous Lows}, where the disk lies in or close to the orbital 
plane. But one would still have to ask, how the free precession can be stable
on long time scales, when it does vary on a time scale of a few years.
2) Or: an intrinsic activity cycle of the optical companion HZ~Her, leading to 
variable mass transfer rates with (quasi) periodic properties 
\citep{Vrtilek_etal01,StillJurua_06,Staubert_etal06a,Staubert_etal09}.
\end{enumerate}

\section{Conclusion}

Based on our analysis of the variations in pulse profiles of Her~X-1 with 
35\,d phase, as observed for a large number of \textsl{Main-Ons} by several
X-ray satellites (with \textsl{RXTE} being the primary source of data), we
conclude that these variations follow the same irregular clock as the flux
turn-ons. There seems to be only one $\sim$35\,d clock in the system which is 
responsible for both, the flux modulation and the variation in pulse profile shape.
The two phenomena are perfectly synchronized and follow each other on 
fairly short time scales.

The important physical question is whether the underlying 35\,d clock is free 
precession of the neutron star which could also be responsible for the pulse profile 
variations. In this case, it needs to be explained how the neutron star can change 
its precessional period every few years by a few percent, and how the neutron star can 
slave the accretion disk so tightly, such that the flux modulation follows immediately.
If there is no free precession of the neutron star, and the clock is due to the accretion
disk alone, then the two most important questions are how the variations in
pulse profile are generated and whether the precession of the accretion disk
can be secularly stable without the existence of an outside force.

 \begin{acknowledgements}
      This paper is based on observational data taken by the NASA satellite
      \textsl{Rossi X-ray Timing Explorer} (RXTE). We like to acknowledge the dedication 
      of all people who have contributed to the great success of this mission.
      D.V. thanks DLR for financial support through grant 50 OR 0702,
      and D.K. acknowledges support by the Carl Zeiss Stiftung.
      We thank the anonymous referee for very valuable comments.
 \end{acknowledgements}

\bibliographystyle{aa}
\vspace{-3mm}
\bibliography{refs_herx1}

\newpage

\appendix
\section{Generation of the pulse profile template}

%Fig. 6
   \begin{figure}
   \includegraphics[angle=-90,width=0.52\textwidth]{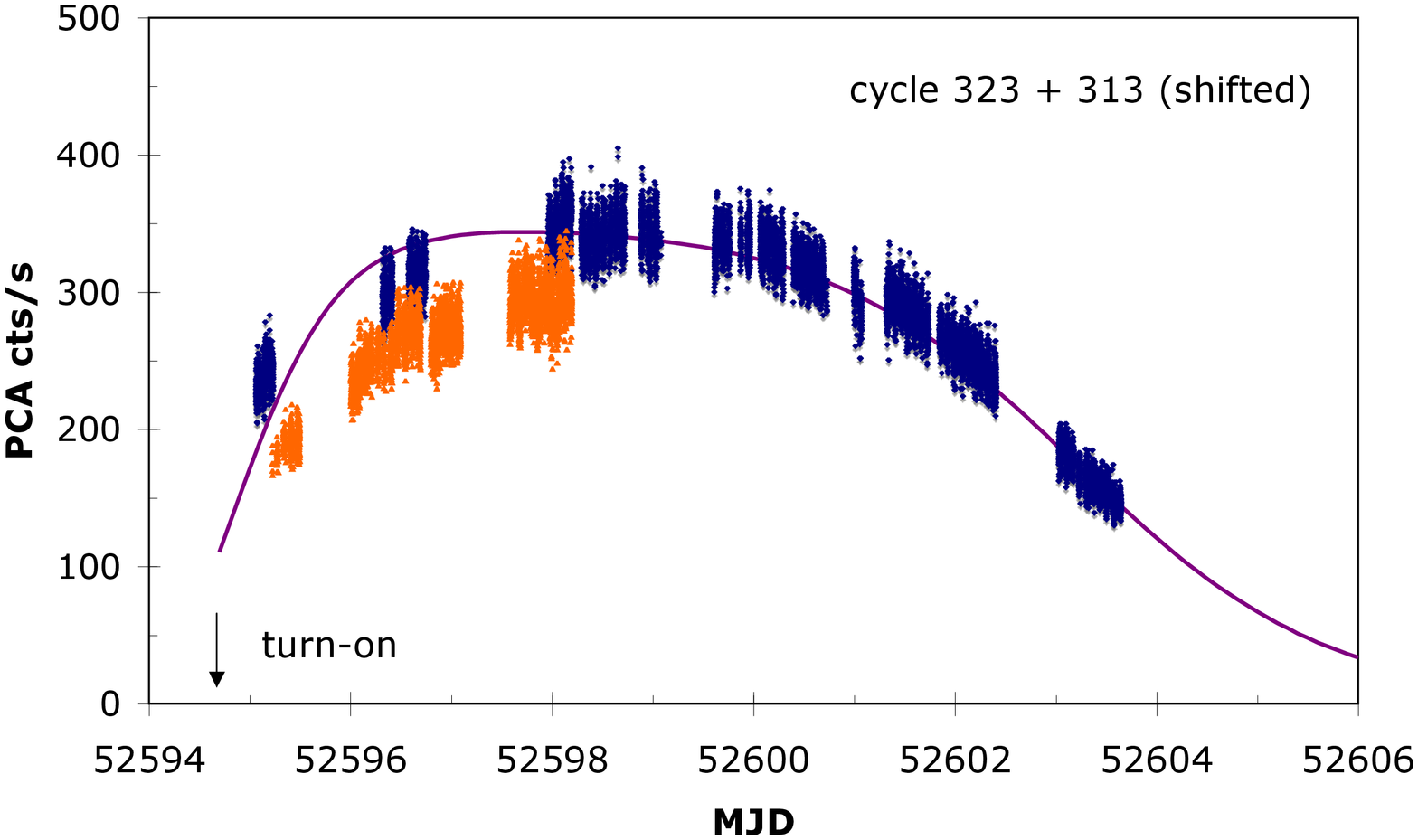}
\vspace*{-7mm}
   \caption{Main-On light curves of 35\,d cycles no. 313 (Dec 01, red) 
   and cycle no. 323 (Nov 02, blue) from \textsl{RXTE}/PCA, 9--13\,keV. 
   The count rate in cycle 313 is $\sim$30\,cts/s lower than in cycle 323,
   here the points are shifted downward by additional 20\,cts/s for separation 
   against the data points of cycle 323. }
   \label{lcs}
\end{figure}

\begin{figure}
\begin{center}
\vspace*{-8mm}
   \includegraphics[angle=-90,width=0.57\textwidth]{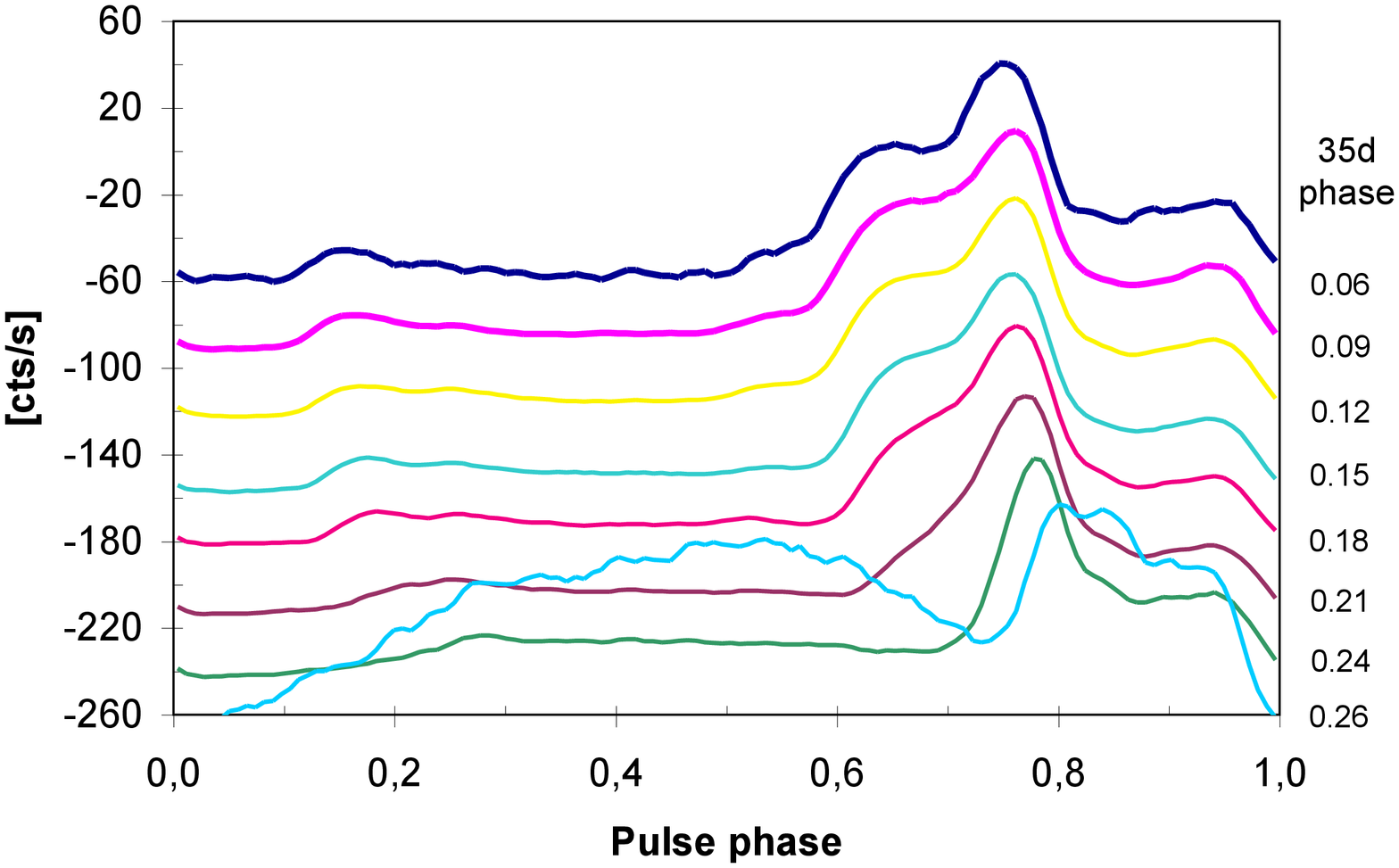}
\vspace{-9mm}
\caption{Selected pulse profiles of Her~X-1 of 35\,day cycles of Dec 01 (cycle no. 313) and Nov 02 
(cycle no. 323) as a function of 35\,day phase (\textsl{RXTE}, 9--13\,keV).}
\label{stacked_obs}
\end{center}
\end{figure}

The X-ray pulse profiles of the 1.24\,s pulsation of Her~X-1 show systematic variations with phase 
of the 35\,day modulation. Here we describe the process of generation of the \textsl{pulse profile 
template} for the \textsl{Main-On} of Her~X-1 in the 9--13\,keV range, its characteristics and the way 
in which the template is used to find the 35\,d phase of any 9--13\,keV pulse profile.
An earlier version of the template (with less resolution) is described in \citet{Staubert_etal10b}. 
To construct the template we use data of the \textsl{Main-Ons} of two 35\,d cycles, taken by 
\textsl{RXTE}/PCA: cycles 313 and 323 of Dec 01 and Nov 02, respectively (for a definition of pulse 
profile cycle counting see Staubert et al., 2009). Together the data provide a near continuous coverage 
of a complete \textsl{Main-On} of Her~X-1. Fig.~\ref{lcs} shows the two light curves in the energy range 
3-20\,keV range (adjusted to a common phase zero).
Fig.~\ref{stacked_obs} shows examples of pulse profiles for progressive 35\,d phases.

\begin{figure*}
   \includegraphics[angle=-90,width=0.53\textwidth]{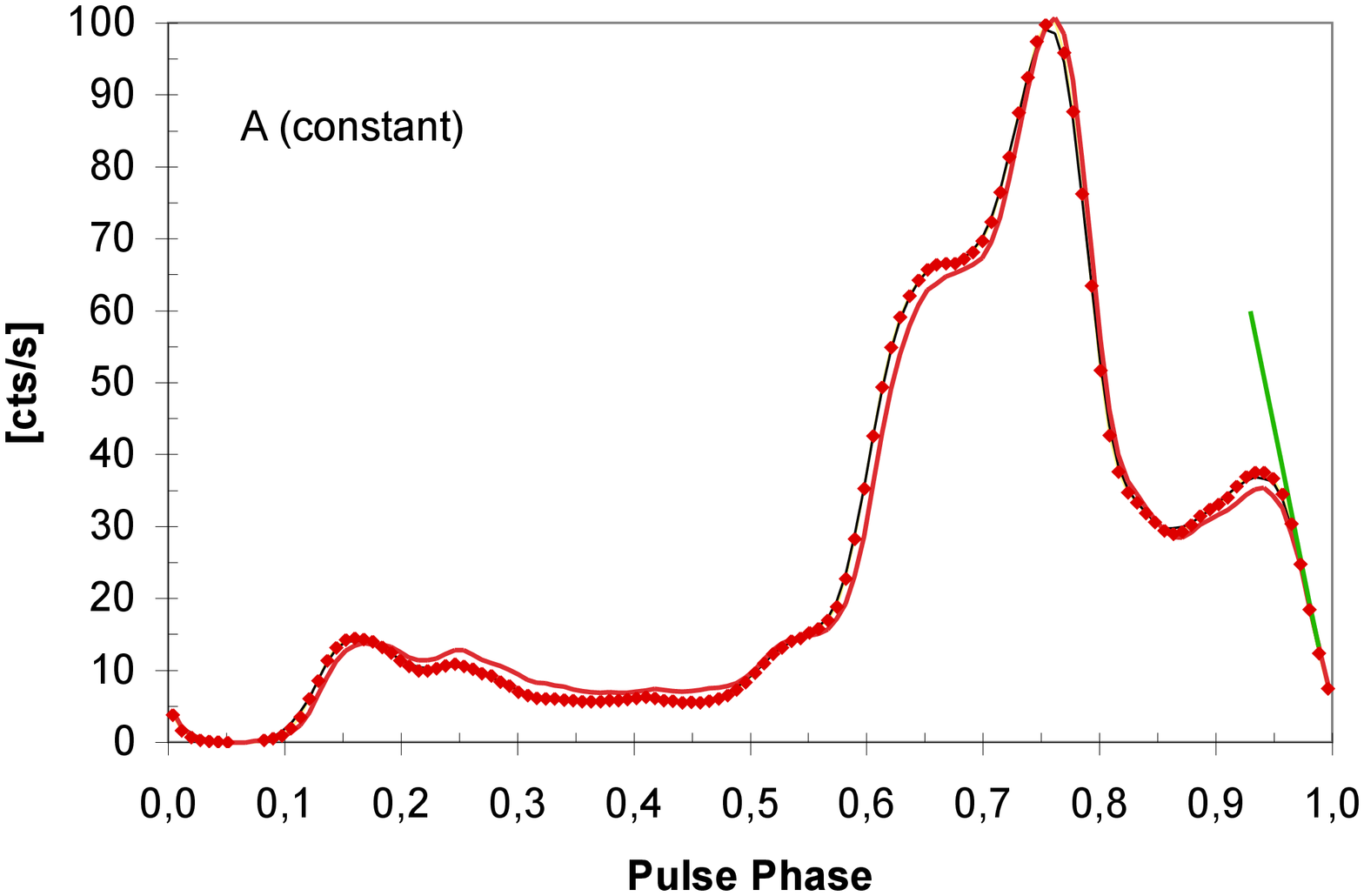} 
   \includegraphics[angle=-90,width=0.53\textwidth]{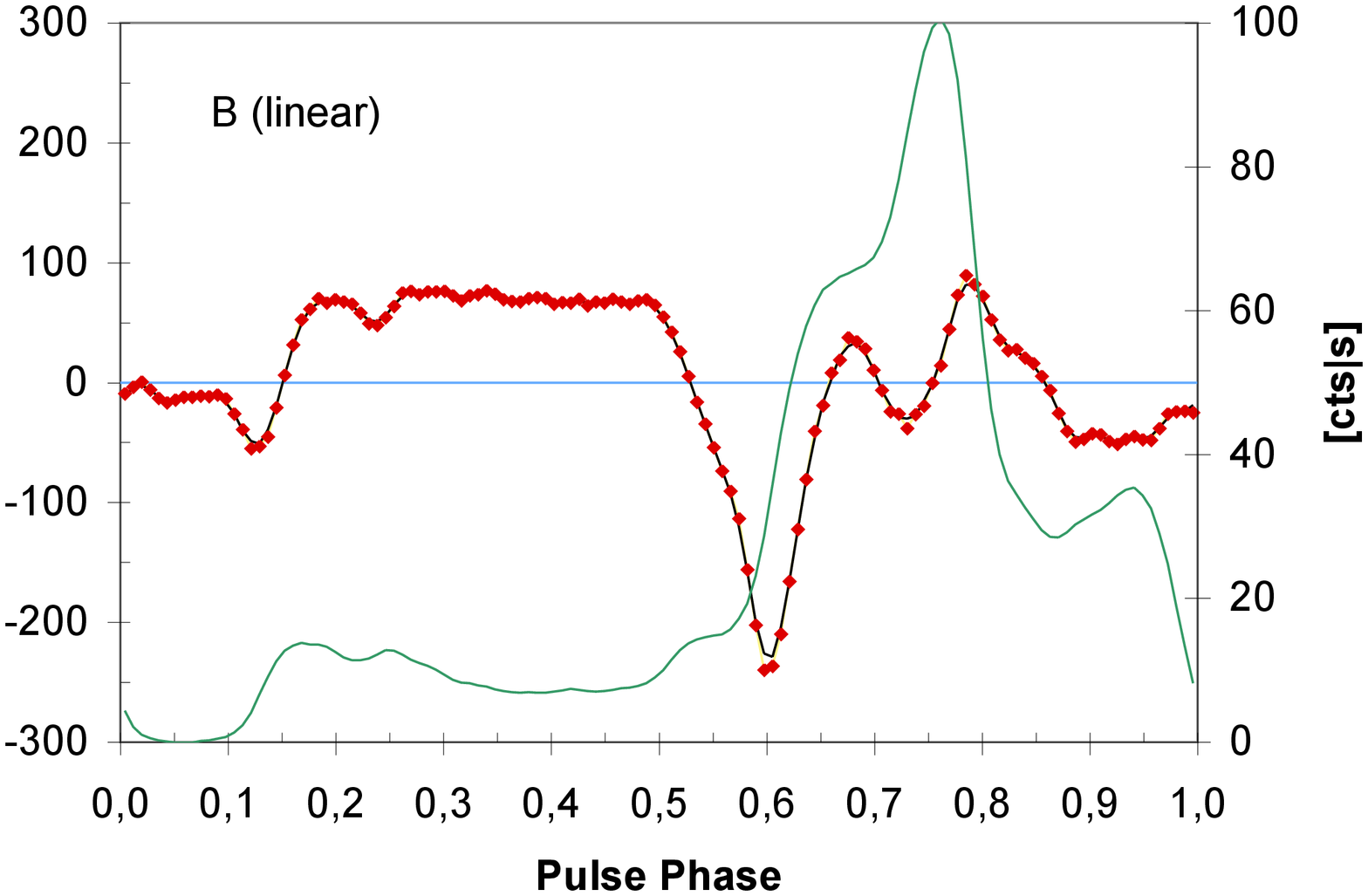}
   \includegraphics[angle=-90,width=0.53\textwidth]{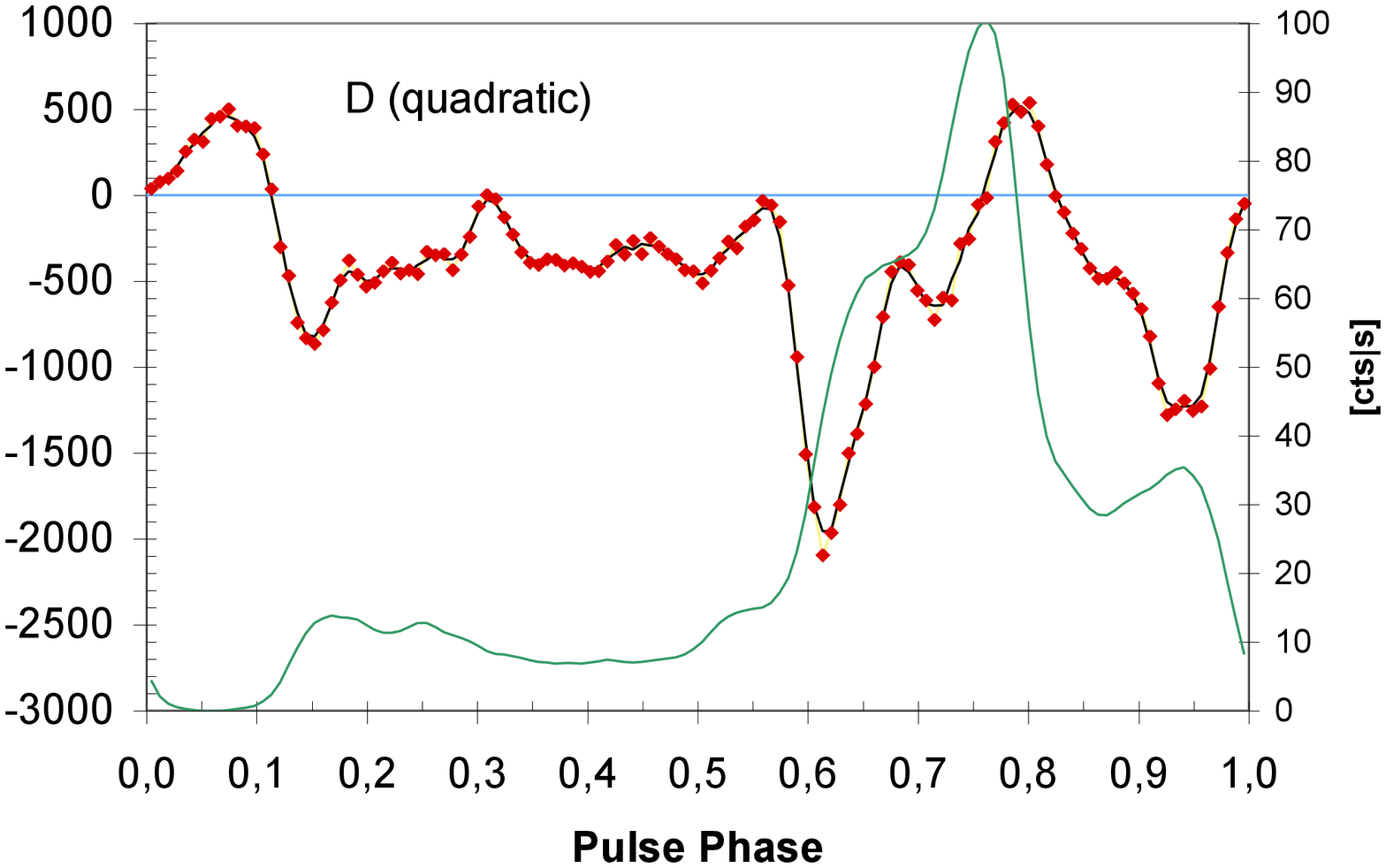}
   \includegraphics[angle=-90,width=0.53\textwidth]{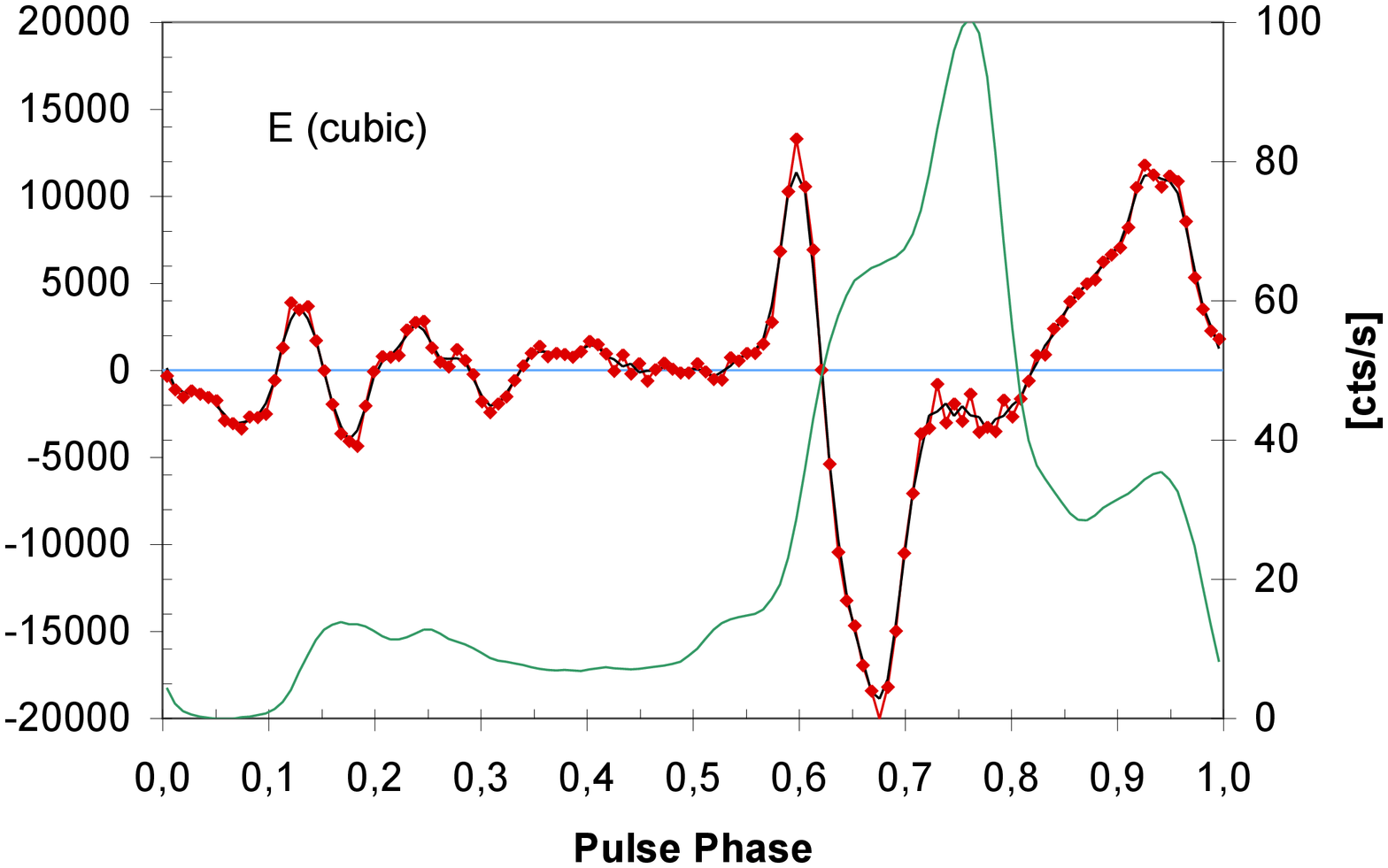}
     \caption{Best fit parameters for cubic fits to the evolution of the count rate
      of normalized pulse profiles for the used 128 pulse phase bins (red points). 
      The reference 35\,d phase is C~=~0.1. Upper left: A(constant), upper right:
      B(linear), lower left: D(quadratic), lower right: E(cubic). 
      For comparison, also the pulse profile for 35\,d phase 0.122 is shown.
      For the parameter A, which is basically the profile itself, the "sharp edge'' at 
      pulse phase 1.0 is drawn.}
   \label{coeff}
\end{figure*}

For generating the template we proceeded through the following steps of analysis:
\vspace*{-1mm}
\begin{enumerate}
\item Generate pulse profiles with 128 phase bins by folding events in the 9--13\,keV range from integration 
         intervals of typically 0.1--0.2\,days ($\sim$0.003--0.006 in 35\,d phase) with the predetermined best pulse period.
         From the available data we find 28 pulse profiles at different 35\,d phases throughout the \textsl{Main-On}. 
\item Normalize all profiles such that the minimum is zero and the maximum is 100\,cts/s.
\item If necessary, shift profiles such that the "sharp edges" of all profiles coincide at pulse phase 1.0. 
         The sharp edges are defined by the fast decrease in flux after the right shoulder of the main peak 
         into the eclipse-like trough (see Staubert et al., 2009b and Fig.~\ref{coeff}-upper-left).
\item Record the 35\,d phase (center of observation interval) for each profile. As 35\,d phase zero the 
         observed turn-ons are used: MJD~52243.05 for cycle no. 313, and MJD~52594.66 for cycle no. 323.
\item Insert all normalized/shifted profiles (each profile is one \textsl{line}) into a matrix according to the 
         corresponding 35\,d phase (which runs along the \textsl{columns}).
\item Perform cubic best fits (see below) with the 35\,d phase as variable (along the \textsl{columns}) for 
         each of the 128 pulse profile bins.
\item Calculate template values using the best fit parameters with a resolution of 0.002 in 35\,d phase.
\end{enumerate}

Cubic fits of the observed normalized count rate values are performed for each of the 128 columns 
using the function \\

%\begin{equation}
\indent F~(cts/s) = A~+~B~$\times$~($\phi$-C)~+~D~$\times$~($\phi$-C)$^{2}$~+~E~$\times$~($\phi$-C)$^{3}$, \\
%\end{equation}

\noindent with $\phi$ being the 35\,d phase (zero phase is \textsl{turn-on}).
Fig.~\ref{coeff} shows the evolution of the four fit parameters with pulse phase. The 35\,d reference 
phase was fixed to C = 0.1. For comparison, also the pulse profile of 35\,d phase 0.122 is shown.

The final 9--13\,keV template is a smoothed analytical representation of the observed pulse profile matrix.
The template covers the \textsl{Main-On} in the 35\,d phase range 0.00 to 0.24, it is a 128~$\times$~121 
matrix of normalized count rate values: each of the 121 lines contains a normalized pulse profile (with 128 bins) corresponding to 35\,d phases ranging from 0.00 to 0.24 in steps of 0.002. A graphical 3D-representation 
of the 9--13\,keV template constructed from the data of 35\,d cycles 313 and 323 is shown in 
Fig.~\ref{compar}-right. The template matrix is electronically provided as an ASCII file.

%Figure 7

%Figure 8

Any 9--13\,keV profile observed during a 35\,d Main-On can then be compared to this 
template and the 35\,d phase can be determined (by $\chi^{2}_{min}$ fitting). 
For the typical \textsl{RXTE} pulse profile, we find that this is generally possible to an accuracy of 
$\pm$0.02 in phase. For the comparison, it is, of course, necessary that the profile which
is to be tested is adjusted in phase such that the "sharp edge'' is at phase 1.0.

In Fig.~\ref{find_phase_zero} we give an example of how "pulse profile phase-zero" is found using
a set of profiles of the \textsl{Main-On} of cycle 257 (July 1996) as observed by \textsl{RXTE}/PCA: 
six different integration intervals were defined for which pulse profiles were generated. The normalized 
profiles were then compared with the analytical template and for each of them the 35\,d phase was 
determined. These phase values are plotted against the observing time and a linear fit with a constant 
slope corresponding to the current 35\,d period value is performed. For cycle 257 the time of 
"pulse profile phase-zero" was found to be MJD 50285.46. This method works with any number of 
observed pulse profiles per Main-On, even with only one profile.

Future work will extend the template to cover also the \textsl{Short-On} of the 35\,d modulation of Her~X-1
and to produce equivalent templates for other energy ranges.

\end{document}